\documentclass[%
 aip,
 amsmath,amssymb,
 reprint,%
]{revtex4-1}

\usepackage{graphicx}% Include figure files
\usepackage{dcolumn}% Align table columns on decimal point
\usepackage{bm}% bold math
\usepackage[utf8]{inputenc}
\usepackage[T1]{fontenc}
\usepackage{mathptmx}
\usepackage{subcaption}
\usepackage{amsmath}
\usepackage[dvipsnames]{xcolor}
\usepackage[version=4]{mhchem}
\usepackage{upgreek}
 
\begin{document}

\preprint{AIP/123-QED}

%\title{Comparative study of atmospheric pressure helium and argon linear plasma jets on methylene blue degradation}
\title{Effect of surface porosity of catalytic supports on plasma-assisted catalysis for ammonia synthesis}
%\author{S. Jaiswal\normalfont\textsuperscript{ b),}}
\author{S. Jaiswal}
%\thanks{Contribute to equally to this study}
\email{surabhijaiswal73@gmail.com}
\affiliation{Princeton University, Department of Chemical and Biological Engineering, Princeton, NJ 08544}
\affiliation{Eastern Michigan University, Department of Physics and Astronomy, 240 Strong Hall, Ypsilanti, Mi 48197}
\author{Zhe Chen}
\thanks{Co-first author and contribute to equally to this study}
\affiliation{Princeton University, Department of Chemical and Biological Engineering, Princeton, NJ 08544}
\author{Sankaran Sundaresan}
\affiliation{Princeton University, Department of Chemical and Biological Engineering, Princeton, NJ 08544}

\author{Bruce E Koel}
\affiliation{Princeton University, Department of Chemical and Biological Engineering, Princeton, NJ 08544}

\email{bkoel@princeton.edu}
\author{Ahmed Diallo}
\affiliation{Princeton Plasma Physics Laboratory, 100 Stellarator Rd, Princeton, NJ 08540}

\begin{abstract}
A fundamental understanding of plasma-catalyst interactions is important for understanding reaction mechanisms, optimizing the catalyst, and increasing the efficiency of plasma-assisted catalytic process for ammonia (\ce{NH3}) synthesis. We report on the effect of the surface porosity of the catalyst support on this reaction carried out in a coaxial dielectric barrier discharge (DBD) plasma reactor. The discharge was created using a variable AC applied voltage at room temperature and near atmospheric pressure (550 Torr). Two catalyst supports were compared: porous silica (\ce{SiO2}) ceramic beads and smooth, non-porous soda lime glass beads of almost equal diameter ($\sim$1.5 mm) were used. \ce{N2} conversion and the \ce{NH3} synthesis rate was increased with increasing voltage for both supports, but the energy yield for \ce{NH3} production increased for the \ce{SiO2} beads and decreased for the glass beads. All three of these parameters were always higher when using the \ce{SiO2} beads, which suggests that porosity can be a small advantage for plasma assisted \ce{NH3} synthesis. Discharge and plasma properties were estimated from Lissajous plots and using calculations with the BOLSIG+ software. The effect of different catalyst supports on the physical properties of the discharge was negligible. High resolution optical emission spectra (OES) were used to explore the evolution of gas phase active species, \ce{N2+}, atomic N, electronically excited \ce{N2}, and atomic H (H$_\alpha$, H$_\beta$), in the plasma in the presence of both supports. The relative concentration of these species was lower in the case of the porous \ce{SiO2} beads for all applied voltages, which suggests that surface reactions are more significant than gas phase reactions for the formation of \ce{NH3} in plasma assisted \ce{NH3} synthesis.

\end{abstract}

\maketitle

\section{\label{sec:intro}Introduction}
%Basic atmospheric plasma
\par The Haber--Bosch (H--B) process is used industrially to produce \ce{NH3} from \ce{N2} and \ce{H2} in a thermal catalytic process at high temperature (600--800 K) and high pressure (100--200 atm).\cite{1} This process is energy intensive and consumes almost 2\% of the world’s annual energy supply.\cite{2} Possible alternative processes to produce \ce{NH3} under milder reaction conditions, such as one utilizing a non-thermal plasma (NTP), are being investigated extensively.\cite{1a, 3} Non-thermal, cold, or non-equilibrium plasmas can produce high energy electrons and create excited molecular species and radicals in the gas phase at ambient conditions of lower temperatures and pressures, enabling reactions with high energy barriers to occur.\cite{4,5} This also enables small scale, fast on/off operations, and possibly reduced energy costs.\cite{2} Therefore, NTP-assisted catalysis for \ce{NH3} synthesis has recently gained increasing attention.\cite{1,4,6,7,8}

\par Investigations using various NTP reactors, including experiments and kinetic modeling, have been performed to understand the fundamentals of plasma-assisted \ce{NH3} synthesis and control mechanisms.\cite{1,4,9,10,11,12,13} Dielectric barrier discharges (DBDs) in different configurations (planar and coaxial) \cite{10,11,12,13} are popular NTP reactors and experiments have been performed with different metal electrodes \cite{1} and dielectric \cite{11} and ferroelectric materials.\cite{12,13} Combining NTPs with heterogeneous catalysts improves the performance for many reactions, including \ce{NH3} synthesis, dry reforming of CH4, and many more.\cite{6,14,15,16} A synergy between the plasma and catalysts is often invoked to explain the enhancement in reaction rate or selectivity, which e.g., enables \ce{NH3} synthesis to proceed at low temperature and pressure conditions.\cite{2} Finding better catalysts and catalyst supports that are optimal in the presence of plasma has been an emerging area of research in recent years. The behavior of plasma in presence of a packed bed of catalysts is also an ongoing area of research.\cite{4}

\par Previous investigations showed that changing the physical properties of the catalyst material such as dielectric constant, surface area, particle size, and void fraction can alter the plasma properties as a result of the modified electric field.\cite{17,18,19,20,21} In addition, the exposures of the plasma to a catalyst can change the chemical or electronic properties of the catalyst, modify surface reaction pathways, change the catalyst morphology, or improving catalyst dispersion.\cite{17,20,22,23} Any or all of these changes may improve the catalyst performance, which is usually quantified as overall reaction conversion and reaction rate.

\par Specifically for plasma-assisted catalysis for \ce{NH3} synthesis, a wide variety of catalysts and catalyst supports have been evaluated.\cite{12,17,21,22,23,24} Patil \textit{et al.} \cite{17} investigated the effect of a range of bare catalyst supports including \ce{\alpha-Al2O3}, \ce{\gamma-Al2O3}, MgO, CaO, \ce{TiO2}, and quartz wool on the synthesis of \ce{NH3} in a DBD reactor. Mehta \textit{et al.} \cite{1} investigated the plasma-assisted catalytic synthesis of \ce{NH3} using a range of \ce{\gamma-Al2O3} supported metal catalysts (Fe, Ru, Co, Ni, and Pt), as well as the bare \ce{\gamma-Al2O3} support particles, in a DBD reactor. They concluded that a key elementary step, the dissociative adsorption of \ce{N2} on metallic catalyst surfaces, can be facilitated by the vibrational excitation of \ce{N2} by energetic electrons in the plasma.\cite{1} The simulation study by Hong \textit{et al.} \cite{9} showed that adsorbed hydrogen H(s) on Fe is essential for plasma-assisted catalytic \ce{NH3} synthesis.

\par In recent years, increasing effort has also been made to understand the plasma behavior, e.g. plasma generation and propagation, in the packed bed reactors that are typically used, where there can be a wide variety of catalysts and  support materials with different shapes, sizes, and pore dimensions. As shown in a two-dimensional simulation of an atmospheric pressure DBD plasma in He, plasma can be generated inside catalyst pores.\cite{21} In later work, Zhang \textit{et al.} \cite{25} modeled the plasma generation in a pore on a dielectric surface. They observed that ionization, electric fields, and electron densities can be larger inside the pores compared to the bulk, and the applied voltage, pore dimensions, and particle shape are critical parameters. Materials of low dielectric constant can have microdischarges in smaller pores, indicating that common catalyst supports such as \ce{\gamma-Al2O3} and \ce{SiO2}, which have low dielectric constants, should have microdischarges in their pores despite the small pore sizes.\cite{21} In another investigation, it was found that at higher voltages, ionization can take place mainly inside the pores.\cite{26} From these investigations, it is clear that the surface porosity of the catalyst support can be an important factor in plasma-assisted catalysis of \ce{NH3} synthesis, and thus it is important to conduct additional studies of the effect of surface porosity in order to advance our fundamental understanding of this reaction process.

\par Recently, a limited number of experiments have been performed by combining optical and electrical diagnostics of the plasma to better characterize the plasma, understand the role of different catalysts, and develop reactions mechanisms based on these measurements.\cite{12,13,24,27} For example, Wang \textit{et al.} \cite{27} performed experiments with Ni, Cu, and Fe catalysts supported on \ce{\gamma-Al2O3} in which they performed various diagnostics for plasma characterization and surface analysis of the materials. However, there has not yet been a report of a study that explored the effect of surface porosity on the overall reaction rate, as well as the plasma characteristics, for plasma-assisted catalysis of \ce{NH3} synthesis.

\par In this paper, we report our results that combine the use of electrical and optical diagnostics of the plasma to measure the discharge characteristics and probe several active gas phase species present in the plasma in a packed-bed coaxial DBD reactor using either porous \ce{SiO2} beads or nonporous soda lime glass (referred to as “glass”) beads and at the same time measured experimentally the \ce{N2} conversion and \ce{NH3} production using gas chromatography (GC).

\begin{figure*}
\includegraphics[width=16.0cm,angle=0]{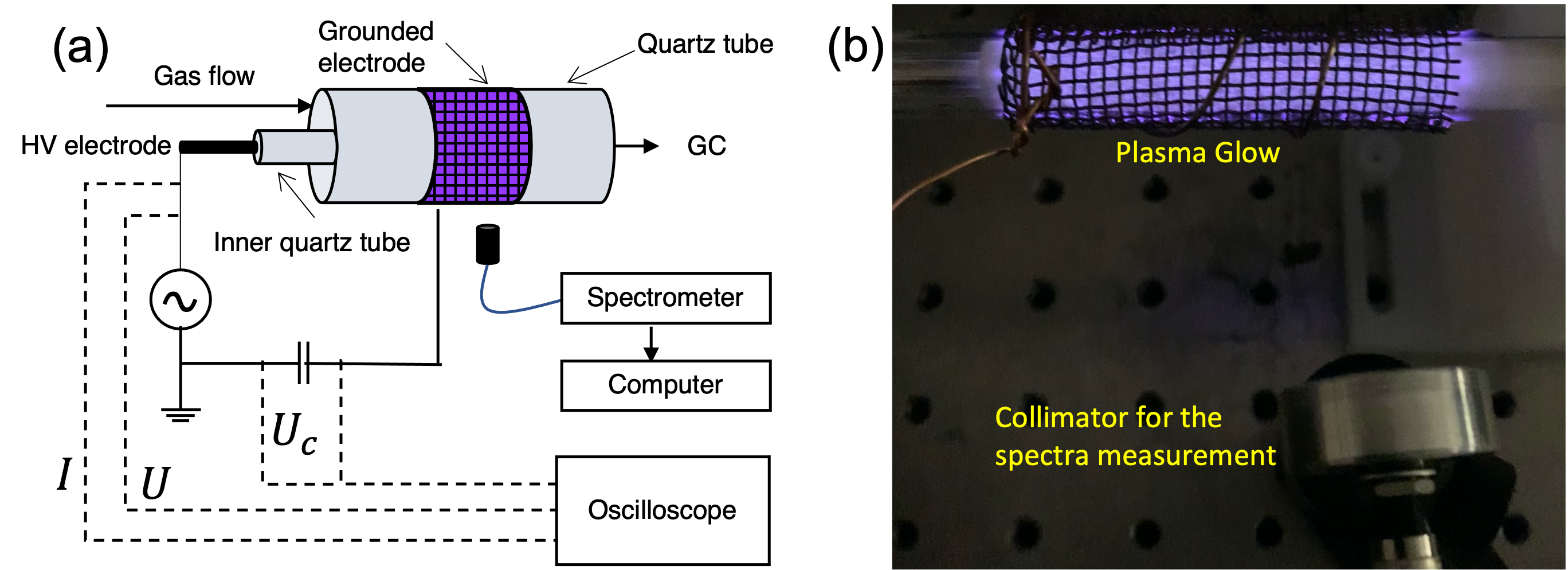}
\caption{\label{fig:schematic} (a) Schematic diagram of the experimental setup. Solid lines and dashed lines represent electrical circuits and electrical signals, respectively. (b) Photograph of the plasma glow in the DBD reactor packed with glass beads and the arrangement of the optical collimator for the OES measurements.}
\end{figure*}

\section{\label{sec:setup}Experimental Setup}
\par Fig.~\ref{fig:schematic}(a) shows a schematic diagram of the experimental setup. The coaxial DBD reactor consisted of a quartz tube (0.4 in. I.D., 0.5 in. O.D.), a 2.25-in. long copper woven wire mesh wrapped around the outside of the quartz tube that served as the grounded electrode, and a 0.125-in. diameter stainless steel rod covered by a 0.25-in. O.D. quartz tube, which served as the high voltage (HV) electrode, running down the center of the larger quartz tube. An AC power source (Information Unlimited, PVM500) with a frequency of 20 kHz was used to produce the DBD plasma.

\subsection{Reaction kinetics experiments}
\par Two different types of dielectric spherical beads, which had the same diameter ($\sim$1.5 mm) but had different surface porosities were used in these experiments: (i) porous \ce{SiO2} (surface area of 300 m$^2$/g) beads, and (ii) non-porous soda lime glass beads. Prior to experiments, the porous \ce{SiO2} beads were dried in-situ at 523 K for 2 h under 50 mL/min \ce{N2}. The glass beads were dried in-situ at 423 K for 2 h under 50 mL/min \ce{N2}. All experiments were run at room temperature, with the reactor pressure set to 550 Torr, and no external reactor heating used. The flow rates of \ce{N2} (Airgas, 99.999\%) and \ce{H2} (Praxair, 99.99\%) into the reactor were controlled by mass flow controllers (Teledyne Hastings Instruments, 300 Series). The total flow rate was kept at 26.5 mL/min. The \ce{NH3} produced in the reactor was quantified by using an online gas chromatograph (Agilent, 7890A) with a CP-Volamine column (Agilent, CP-7447) and a thermal conductivity detector.

\subsection{Plasma characterization}
\par Signals of applied voltage, current, and the voltage across an external capacitor (1000 pF) connected in series with the reactor were recorded by an oscilloscope (GW Instek, GDS-1054B). A Lissajous plot, which is a plot of charge vs. applied voltage, was constructed from the signals of the capacitor voltage and applied voltage.\cite{28} Discharge power was estimated from the Lissajous plot as the product of frequency and the area enclosed by the plot.\cite{28} We then used BOLSIG+ \cite{31} to estimate the mean electron temperature of the plasma using the reduced electric field as an input, which was estimated from the Lissajous plot following Butterworth \textit{et al.} \cite{29} and Mei \textit{et al.} \cite{30} We also estimated the electron number density from the Lissajous plot using a simple technique described by Hong \textit{et al.} \cite{32}, in which the electron mobility was obtained by using BOLSIG+.\cite{31} Optical emission spectra were recorded using a high resolution ($\leq$0.05 nm) spectrometer (Princeton Instruments SpectraPro$\textsuperscript{\textregistered}$ HRS-500). The optical arrangement consisted of a multiwavelength (200–1000 nm) collimator (Fig.~\ref{fig:schematic}(b)) that is connected to the spectrometer via a 1-m long optical fiber. The optical setup was located at a distance of $\sim$10 cm from the reactor. A broad spectrum was measured by using a 1200 g/mm grating in the Step and Glue mode that enabled covering a wavelength range from 300 - 1100 nm without compromising the resolution. The spectrometer integration time was set to between 1 - 15 s to compensate for the varying signal from the plasma.

 \begin{figure*}
\includegraphics[width=17.0cm,angle=0]{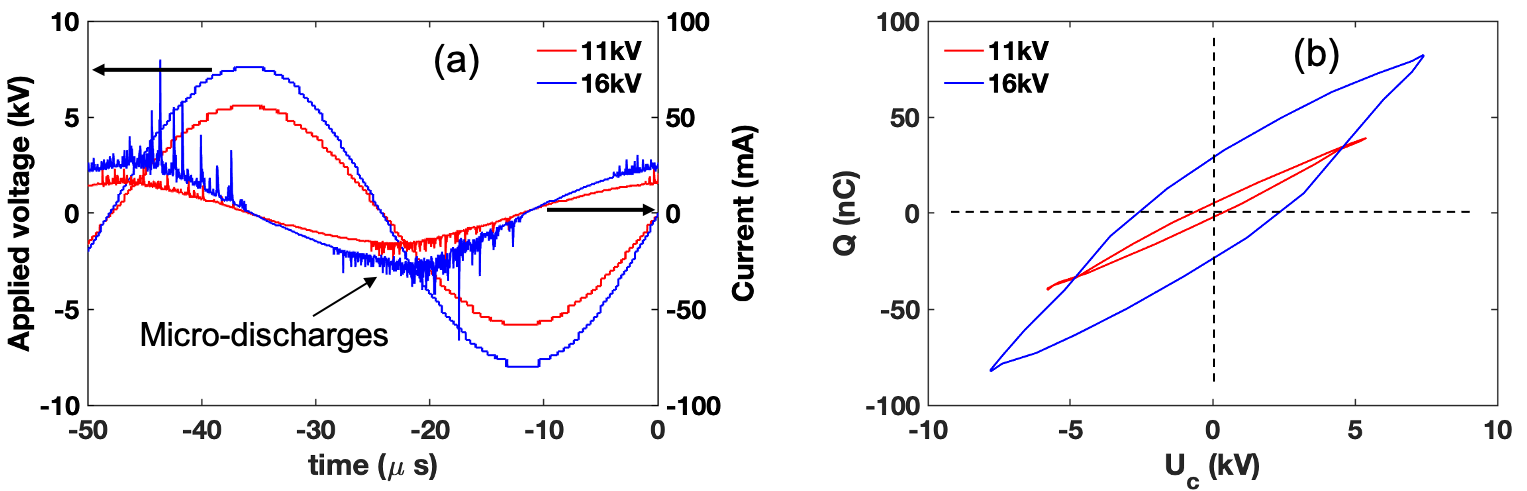}
\caption{\label{fig:fig2}(a) Applied voltage and current signals and (b) the corresponding Lissajous plots for the DBD reactor packed with glass beads at an applied voltage of 11 and 16 kV. Conditions: 550 Torr, 26.5 mL/min, \ce{N2}:\ce{H2}=1:3.}
\end{figure*}

\par Measured signals of both applied voltage and current for the reactor packed with glass beads at 11 and 16 kV are shown in Fig.~\ref{fig:fig2}(a). Both the number and the intensity of microdischarges increased with applied voltage. Lissajous plots obtained using a packed bed of glass beads for applied voltage 11 and 16 kV are shown by Fig.~\ref{fig:fig2}(b). As the applied voltage was increased, the area enclosed by the plot increased, indicating that the consumed power increased. Similarly, the reduced electric field increased with applied voltage, and the values of the reduced electric field at various applied voltages are shown in Table S1 of the Supplementary Information (SI).

\section{\label{sec:discussion}Results and Discussion}
\subsection{Materials characterization}
\par
\begin{figure*}
%\begin{subfigure}{0.47\textwidth}
\includegraphics[width=15.0cm,angle=0]{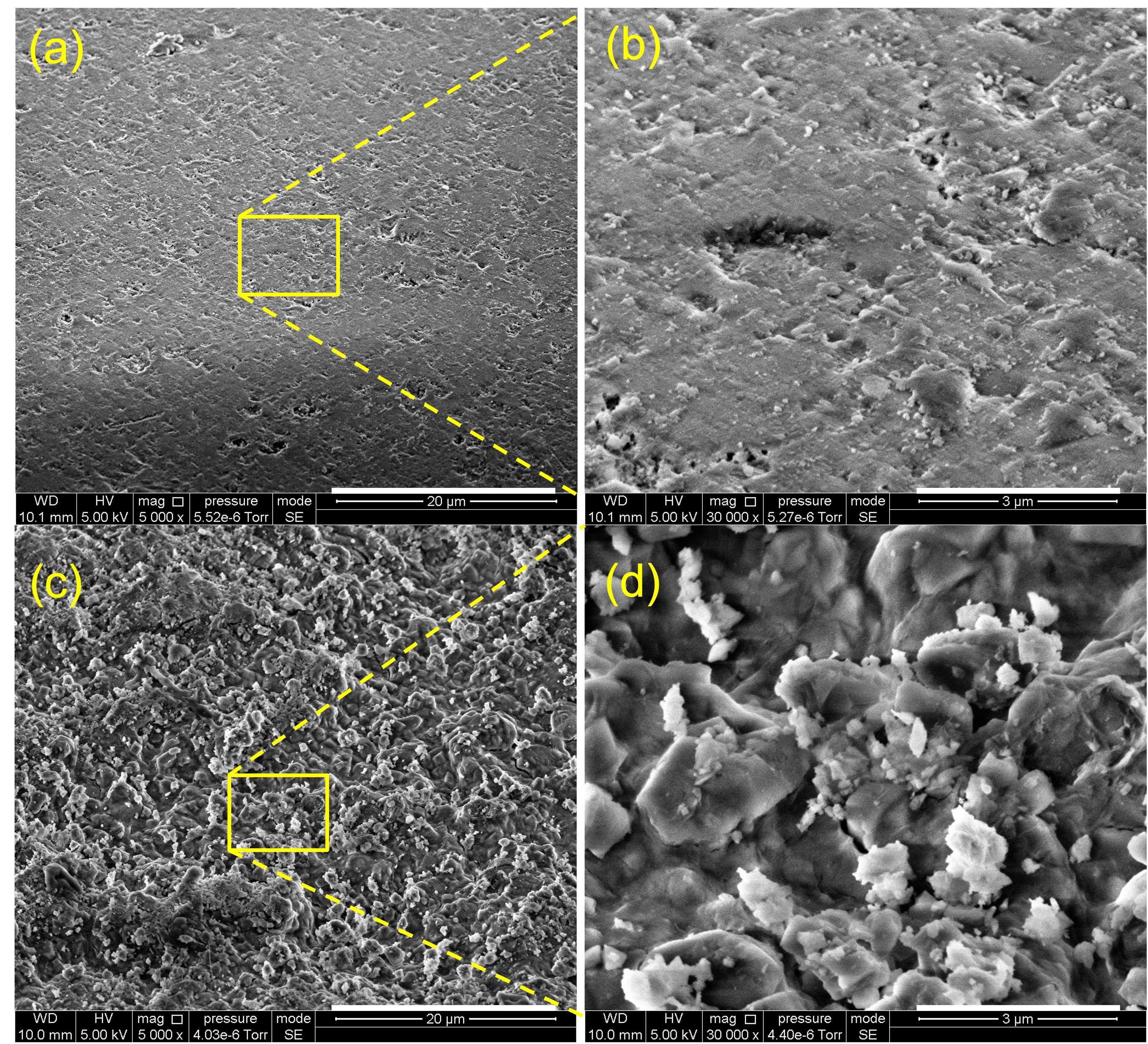}
\caption{\label{fig:fig3} SEM images of the surface of a nonporous glass bead (a and b) and a porous \ce{SiO2} bead (c and d). Scale bars are 20 $\upmu$m for (a) and (c) and 3 $\upmu$m for (b) and (d).}
\end{figure*}

 Surface area and pore size distribution of the \ce{SiO2} beads was measured using a Brunauer--Emmett--Teller (BET) analyzer (Micromeritics 3Flex). The \ce{SiO2} surface area was 300 m$^2$/g and the average pore size was 8 nm. The glass beads were nonporous with a calculated geometric surface area of $3.4\times10^{-3}$ m$^2$/g assuming 2.1 as the surface roughness.\cite{9} SEM images were taken for both the \ce{SiO2} and glass beads for additional characterization, and these are shown in Fig.~\ref{fig:fig3}. For better visualization, a zoomed-in view of each of the respective images in Figs.~\ref{fig:fig3} (a) and (c) (marked by yellow rectangles) are shown in Figs.~\ref{fig:fig3} (b) and (d). The surfaces of the nanoporous \ce{SiO2} beads also have a microscale roughness much larger than that of the nonporous glass beads.

%%%%%%%%%%%%%%%%%%%%%%%%%%%%
\subsection{Effects of surface porosity and applied voltage}

\begin{figure}
\includegraphics[width=8.0cm,angle=0]{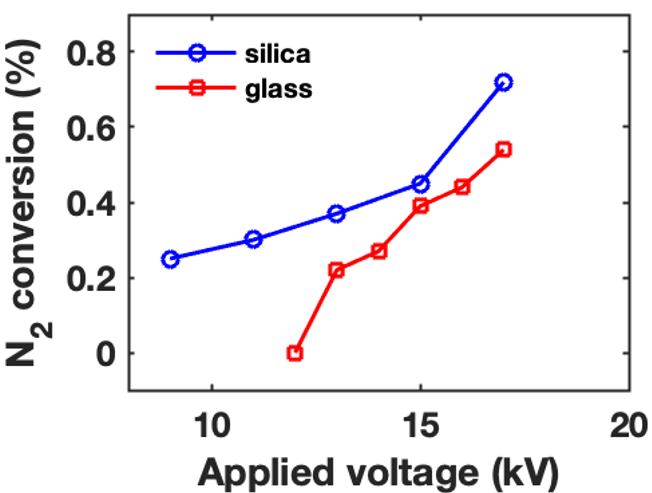}
\caption{\label{fig:fig4} Observed \ce{N2} conversion for different applied voltages for the DBD reactor packed with \ce{SiO2} or glass beads. Reaction conditions: 550 Torr, 26.5 mL/min, \ce{N2}:\ce{H2}=1:3.}
\end{figure}
%%%%%%%%%%%%%%%%%
\begin{figure}
\includegraphics[width=8.0cm,angle=0]{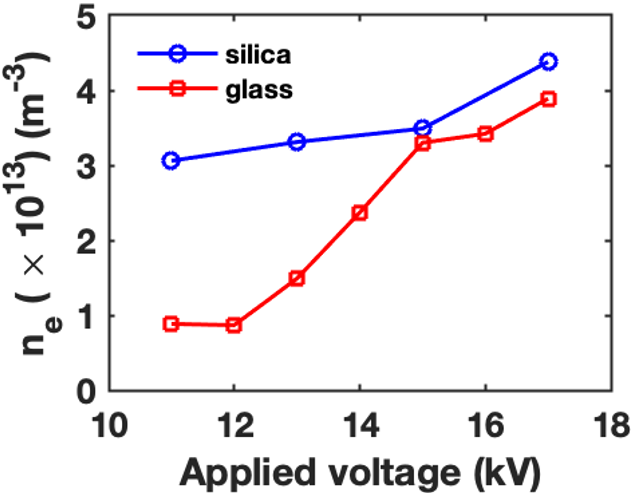}
\caption{\label{fig:fig5} Electron number density determined from Lissajous plots for different applied voltages for the DBD reactor packed with \ce{SiO2} or glass beads. Reaction conditions: 550 Torr, 26.5 mL/min, \ce{N2}:\ce{H2}=1:3.}
\end{figure}

\par Experiments were conducted using a feed ratio of \ce{N2}:\ce{H2} = 1:3 at various applied voltages for the discharge. \ce{N2} conversion was calculated from the \ce{NH3} mole fraction measured by the GC under each experimental condition. As shown in Fig.~\ref{fig:fig4}, \ce{N2} conversion increased with increased applied voltage for both \ce{SiO2} and glass beads, with the \ce{SiO2} beads producing higher \ce{N2} conversion compared to the glass beads. Electron number densities were estimated from the Lissajous plots measured for each experiment. Fig.~\ref{fig:fig5} shows that a higher applied voltage resulted in a higher electron number density, although this effect is much less for packing with porous \ce{SiO2} beads. Surface porosity of the packing beads clearly has an influence on the electron number density, such that at the same applied voltage, the electron number density was higher when using the porous \ce{SiO2} beads than with the nonporous glass beads, however, the difference in the electron densities is of little significance at the higher applied voltages. 

\begin{figure*}
\includegraphics[width=16.0cm,angle=0]{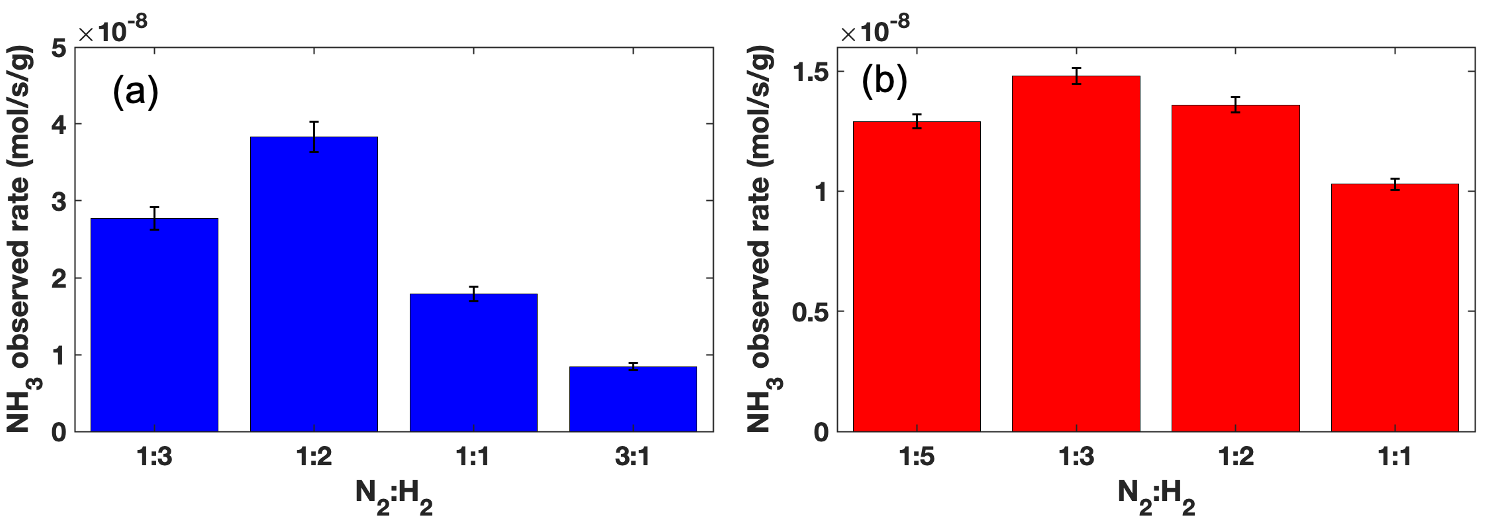}
\caption{\label{fig:fig6} \ce{NH3} observed production rate at different \ce{N2}:\ce{H2} feed ratios for the DBD reactor packed with (a) \ce{SiO2} and (b) glass beads. Reaction conditions: 550 Torr, 26.5 mL/min, 17 kV.}
\end{figure*} 
 \begin{figure*}
\includegraphics[width=16.0cm,angle=0]{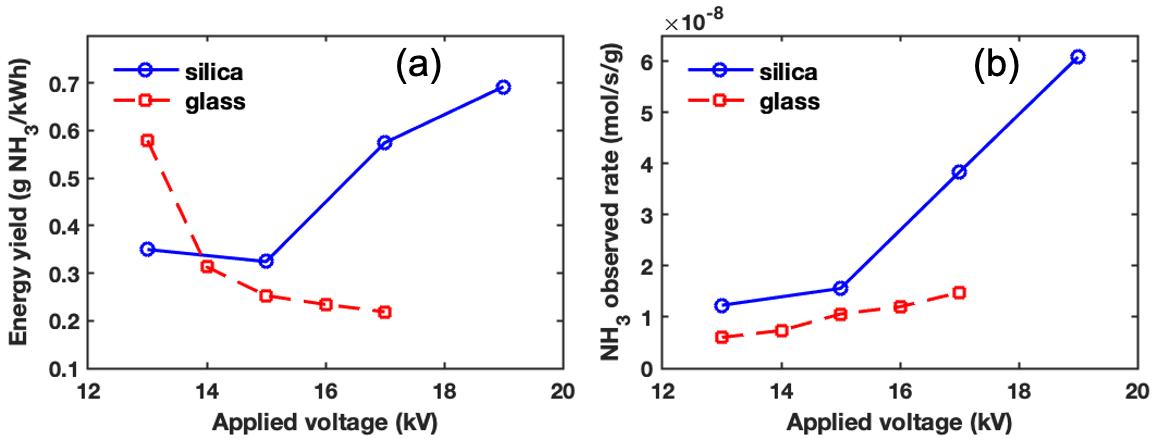}
\caption{\label{fig:fig7} (a) \ce{NH3} energy yield and (b) \ce{NH3} observed production rate at optimal \ce{N2}:\ce{H2} ratio (1:2 for \ce{SiO2} and 1:3 for glass beads) for different applied voltages. Reaction conditions: 550 Torr, 26.5 mL/min.}
\end{figure*}

\par It was reported that the \ce{N2}:\ce{H2} feed ratio had an effect on the observed rate of \ce{NH3} production.\cite{1,12} For example, when using \ce{\gamma-Al2O3} and \ce{Ru/\gamma-Al2O3} catalysts in the reactor, the optimal ratio of \ce{N2}:\ce{H2} was found to lie between 1:1 and 3:1, richer in \ce{N2} compared to the stoichiometric ratio.\cite{1} Thus, it is important to also compare the reaction rates and energy yields in the presence of \ce{SiO2} and glass beads at their respective \ce{N2}:\ce{H2} optimal feed ratios. We first performed \ce{NH3} synthesis experiments with both catalyst support packing materials using various feed ratios to determine the \ce{N2}:\ce{H2} ratio that gave the highest reaction rate. As shown in Fig.~\ref{fig:fig6}, with \ce{SiO2} beads, the optimal feed ratio was 1:2. This indicates that a nitrogen-rich feed resulted in a higher observed \ce{NH3} production rate. However, further increases in the \ce{N2}:\ce{H2} ratio reduced the reaction rate. With glass beads, the optimal \ce{N2}:\ce{H2} ratio was 1:3, which represents a less nitrogen-rich feed from that using \ce{SiO2} beads.

\par Next, we compared the \ce{NH3} energy yields for the DBD reactor packed with \ce{SiO2} and glass beads, as well as the observed \ce{NH3} production rates of these two catalyst support materials at their respective optimal feed ratios (\ce{N2}:\ce{H2}=1:2 for \ce{SiO2} and 1:3 for glass beads). As shown in Fig.~\ref{fig:fig7}, packing with \ce{SiO2} beads gave a higher \ce{NH3} energy yield than that when using glass beads except at the lowest applied voltage tested at 13 kV. A higher observed \ce{NH3} production rate was also seen for packing with \ce{SiO2} beads compared to that when using glass beads at all of the voltages tested in our experiments. With increased applied voltage, the \ce{NH3} energy yield increased when using \ce{SiO2} beads, but decreased for glass beads, and the observed \ce{NH3} production rate increased more significantly in the presence of \ce{SiO2} beads compared to glass beads. This may be explained by the generation of more surface microdischarges at higher voltages due to the porous surface structure of \ce{SiO2} compared to that of the glass beads.

%%%%%%%%%%%%%%%%%%%%%%%%%%%%%%%
\begin{figure*}
\includegraphics[width=17.0cm,angle=0]{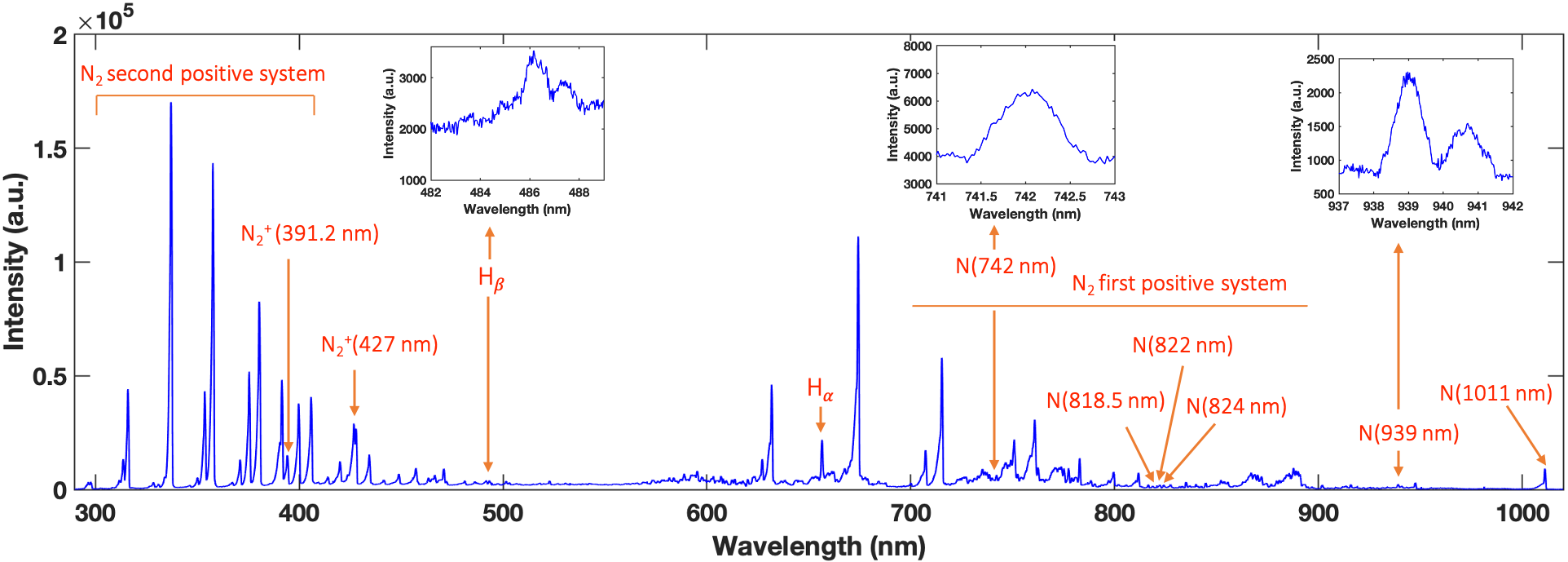}
\caption{\label{fig:fig8} OES spectrum of the plasma during operation of the DBD reactor packed with glass beads at an applied voltage of 16 kV. Conditions: 550 Torr, 26.5 mL/min, \ce{N2}:\ce{H2}=1:3. }
\end{figure*}

\subsection{OES spectra and intensities of gas phase species}
\label{OES}
\par
\begin{figure*}
\includegraphics[width=14.0cm,angle=0]{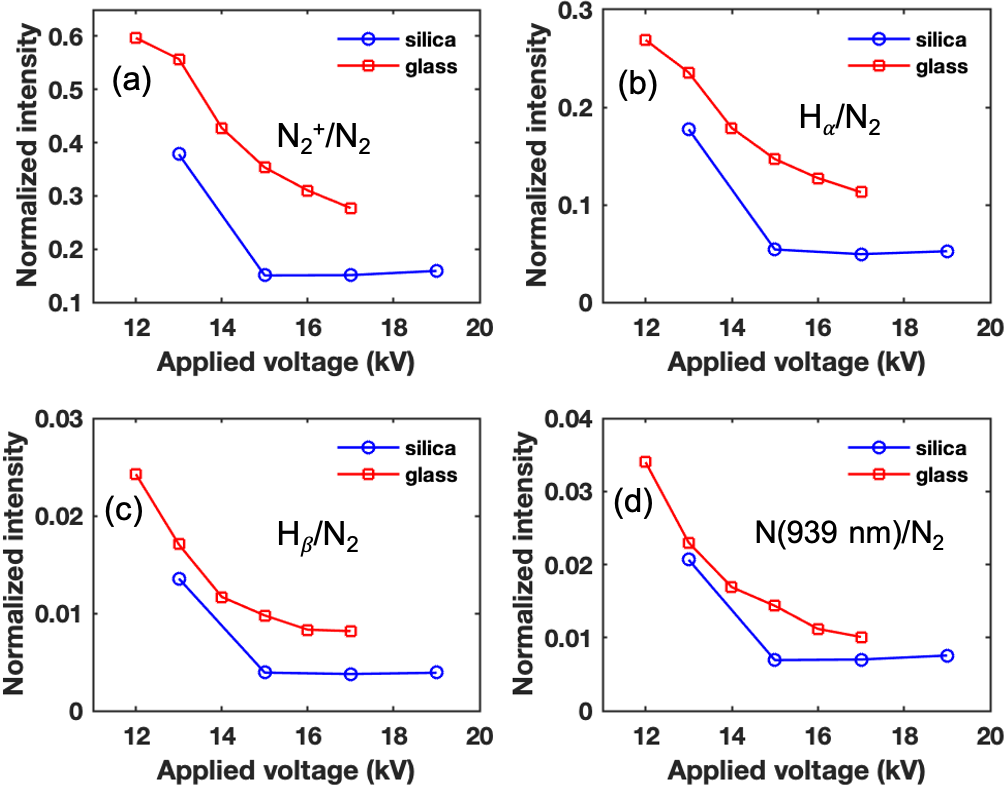}
\caption{\label{fig:fig9} Normalized intensities of N${_2}^+$, H$_\alpha$, H$_\beta$, and N (939 nm) emission lines from the plasma during operation of the DBD reactor packed with \ce{SiO2} or glass beads. Conditions: 550 Torr, 26.5 mL/min, \ce{N2}:\ce{H2}=1:3.}
\end{figure*}

 OES spectra of the plasma were measured during operation of the DBD reactor packed with \ce{SiO2} or glass beads at several applied voltages. Fig.~\ref{fig:fig8} shows a representative OES spectrum from the DBD reactor packed with glass beads with an applied voltage of 16 kV as an example. Observation of the \ce{N2} second positive system and the \ce{N2} first positive system suggests that \ce{N2} molecules are electronically excited by electrons in the plasma: \ce{e + N2 -> e + N2^*}. Observation of the \ce{N2+} first negative system at 391.2 nm and 427 nm confirmed electron-impact ionization in the plasma: \ce{e + N2 -> 2e + N2+}. The hydrogen Balmer lines, with H$_\beta$ at 486.2 nm and H$_\alpha$ at 656 nm, indicates that atomic H was present in the plasma. The existence of atomic N species in the plasma was confirmed by emission lines at several wavelengths: 742, 818.5, 822, 824, 939, and 1011 nm. Identification of H Balmer atomic lines and atomic N suggests the dissociation of \ce{N2} and \ce{H2}.\cite{27}  We did not clearly observe an NH line at 336 nm, but this might overlap with the 337-nm line of the \ce{N2} second positive system. The OES spectra provide direct evidence of the existence of several gas phase active species in the \ce{N2}-\ce{H2} plasma during \ce{NH3} synthesis such as electronically excited molecular nitrogen \ce{N2^*}, \ce{N2+}, N, and H. The spectrum shown here is mostly consistent with previous reports \cite{12,27} on \ce{N2}-\ce{H2} plasma generated in a DBD reactor with different packing materials. However, those other reported spectra were not well resolved and so the measurement of an NH line at 336 nm was obtained by using the shoulder of the \ce{N2} second positive system at 337 nm. We did not detect any peak at 336 nm even at very high resolution (0.05 nm) measured using a 2400 g/mm grating, and therefore, we do not assign an NH line in our spectra. The normalized intensities of emission lines for \ce{N2+}, H$_\alpha$ , H$_\beta$ , and N (939 nm) (relative to the intensity of the \ce{N2} line at 357 nm \cite{12}) for the plasma during operation of the DBD reactor packed with \ce{SiO2} and glass beads are shown in Figs.~\ref{fig:fig9}(a)-(d) for several applied voltages. For both packing materials, the relative intensities of all of the emission lines decreased with increasing applied voltage. Assuming that the ratio between band intensities is equivalent to that between the concentrations \cite{12} of the emitting species, we can see that increasing applied voltage increased the plasma energy that was expended in  electronically exciting \ce{N2} molecules more than in generating \ce{N2+}, H, and N through electron-impact ionization and electron-impact dissociation. Furthermore, as shown in Fig.~\ref{fig:fig9}, relative intensities of \ce{N2+}, H$_\alpha$ , and atomic nitrogen lines were found to be higher with glass bead packing than with \ce{SiO2}. This is because at the same applied voltage, the plasma had a lower reduced electric field and hence a lower mean electron temperature for \ce{SiO2} relative to glass packing. The values for the reduced electric field and mean electron energy for these experiments are provided in the SI. Because the rate constants for electron-impact ionization (\ce{e + N2 -> 2e + N2+}) and electron-impact dissociation (\ce{e + N2 -> e + 2N} and \ce{e + H2 -> e + 2H}) increase with mean electron temperature \cite{33}, these reactions proceed at lower reaction rates in the presence of \ce{SiO2} compared to glass beads and lead to fewer \ce{N2+}, H, and N species in the plasma.

%%%%%%%%%%%%%%%%%%%%%%
%%%%%%%%%%%%%%%%%%%%%%%%%%%%%%%
 \begin{figure}
\includegraphics[width=8.5cm,angle=0]{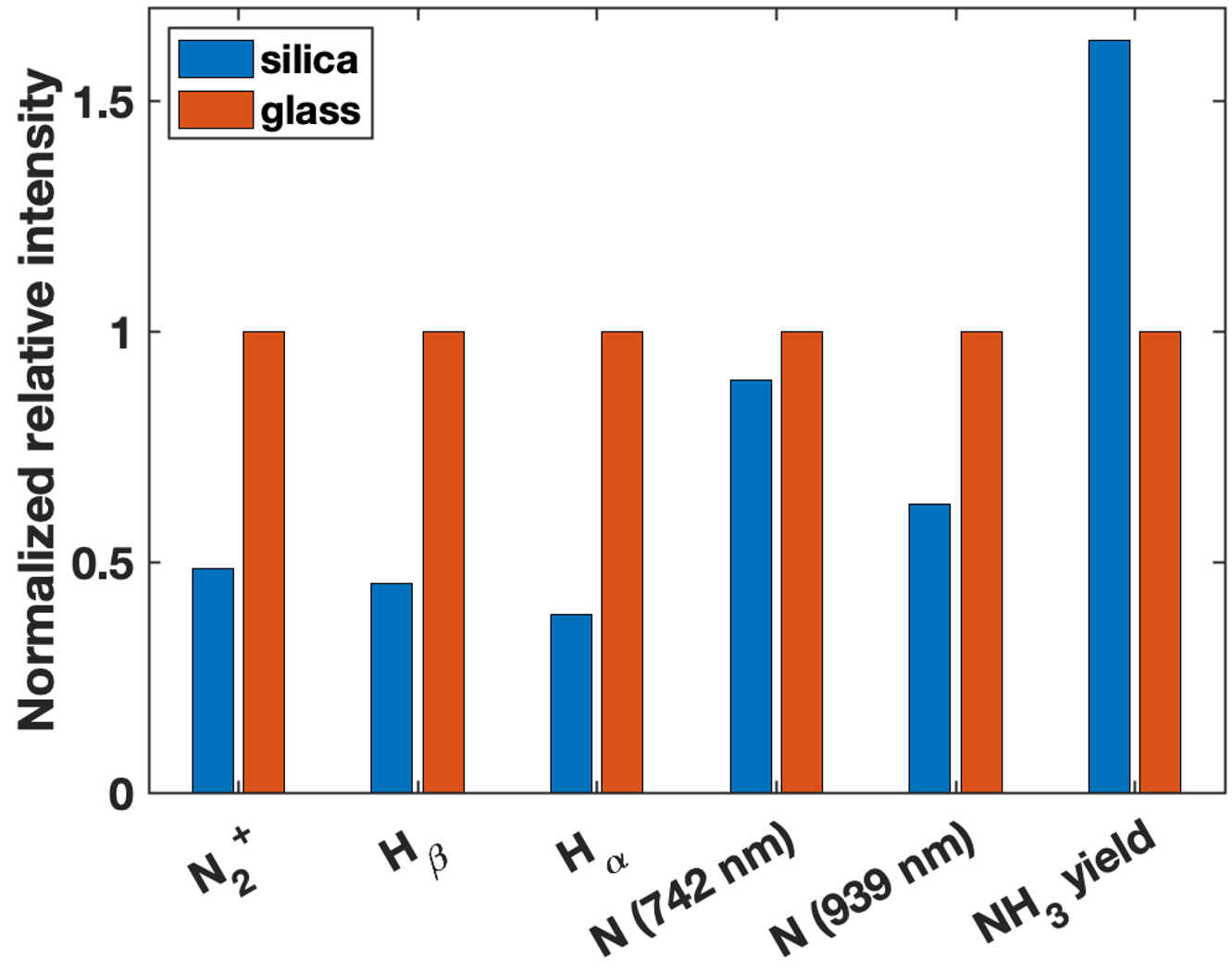}
\caption{\label{fig:fig10} Normalized relative intensities of N${_2}^+$, H$_\beta$, H$_\alpha$, N(742 nm), and N(939 nm) emission lines from the plasma during operation of the DBD reactor packed with \ce{SiO2} and glass beads at 11 W. Conditions: 550 Torr, 26.5 mL/min, \ce{N2}:\ce{H2}=1:3, and an applied voltage of 17 kV for \ce{SiO2} and 16 kV for glass beads.}
\end{figure}
%%%%%%%%%%%%%%%%%%%%%%

\par We also compared the normalized intensities of \ce{N2+}, H$_\beta$, H$_\alpha$, and N lines at 742 nm and 939 nm for the DBD reactor packed with \ce{SiO2} and glass beads at the same plasma power of 11 W, corresponding to an applied voltage of 17 kV for \ce{SiO2} and 16 kV for glass beads. Fig.~\ref{fig:fig10} shows the normalized relative intensities of these lines, where the intensities were ratioed against the N(357 nm) line and then normalized with respect to those intensities for glass (set to unity) following the same approach as that by Wang \textit{et al.}\cite{27} The \ce{NH3} yield is also shown on the right of Fig.~\ref{fig:fig10},  which was normalized with respect to the yield obtained with glass bead packing (set to unity). At the same plasma power, the intensities of the five emission lines monitored were found to be lower in the presence of \ce{SiO2} than glass beads. As noted before, this could be due to a lower mean electron energy with \ce{SiO2} packing leads to lower concentrations of the related gas phase species. Interestingly, the observed \ce{NH3} yield with \ce{SiO2} packing was found to be more than 1.5 times higher than that with glass packing even though \ce{SiO2} packing had lower concentrations of active plasma species such as \ce{N2+}, H, and N. 

\par As concluded by Hong \textit{et al.} \cite{9} in their zero-dimensional kinetic modelling results, radical adsorption and Eley--Rideal (E--R) surface reactions can occur on non-metallic surfaces such as \ce{Al2O3}, as given below.
\newline
\newline
Radical adsorption: 
\begin{equation}
    \ce{N + surf -> N(s)}
\label{rxn:N+surf}
\end{equation}
\begin{equation}
    \ce{H + surf -> H(s)}
\label{rxn:H+surf}
\end{equation}
\begin{equation}
    \ce{NH + surf -> NH(s)}
\label{rxn:NH+surf}
\end{equation}
\begin{equation}
    \ce{NH2 + surf -> NH2(s)}
\label{rxn:NH2+surf}
\end{equation}
\newline
E--R reactions: 
\begin{equation}
    \ce{N + H(s) -> NH(s)}
\label{rxn:N+H(s)}
\end{equation}
\begin{equation}
    \ce{NH + H(s) -> NH2(s)}
\label{rxn:NH+H(s)}
\end{equation}
\begin{equation}
    \ce{NH2 + H(s) -> NH3}
\label{rxn:NH2+H(s)}
\end{equation}
\begin{equation}
    \ce{H + N(s) -> NH(s)}
\label{rxn:H+N(s)}
\end{equation}
\begin{equation}
    \ce{H + NH(s) -> NH2(s)}
\label{rxn:H+NH(s)}
\end{equation}
\begin{equation}
    \ce{H + NH2(s) -> NH3 + surf}
\label{rxn:H+NH2(s)}
\end{equation}

The rate constants for these surface reactions can be calculated using the following equation \cite{9}: 
\begin{equation}
    k = \left[\frac{\Lambda}{D} + \frac{V}{A}\frac{2(2-\gamma)}{\overline{v} \gamma}\right]^{-1} {S_\textup{T}}^{-1}
\label{eqn:rate_const}
\end{equation}
where $\Lambda$ is the diffusion length, $D$ is the diffusion coefficient, $V$ is the discharge volume, $A$ is the surface area, $\gamma$ is the sticking coefficient, $\overline{v}$ is the thermal velocity, and $S_\textup{T}$ is the total surface site density. As shown by Eq.~(\ref{eqn:rate_const}), rate constants for radical adsorption and E--R reactions increase with increasing surface area. As a result, \ce{SiO2} beads have higher surface area available for radical adsorption and E--R reactions compared to nonporous glass beads, which results in higher rate constants for these surface reactions. 

\par Although the OES data show that concentrations of N and H were lower in the presence of \ce{SiO2} beads than glass beads, a higher \ce{NH3} observed reaction rate and a higher \ce{NH3} yield were obtained with \ce{SiO2} packing, suggesting that these species must be consumed more in the \ce{SiO2} bed. This is likely due to the fact that the higher surface porosity of \ce{SiO2} provided larger surface area, which resulted in higher rate constants for surface reactions in Eqns.~(\ref{rxn:N+surf}--\ref{rxn:H+NH2(s)}). The simulation study by Hong \textit{et al.} showed that radical adsorption and E--R reactions can proceed on non-metallic surfaces and provide a pathway to \ce{NH3} even in the absence of catalytic metal nanoparticles.\cite{9} Thus, our operando studies combining OES and GC data together indicate that surface porosity had important effects on the overall reaction and surface chemistry had a more significant contribution to the production of \ce{NH3} than gas phase chemistry in these porosity studies.

\section{\label{sec:con}Conclusion}
\par We have investigated the effect of surface porosity of catalytic supports on the reaction rate and energy yield of plasma-assisted catalysis for \ce{NH3} synthesis. We performed \ce{NH3} synthesis experiments using a coaxial DBD reactor packed with two different catalyst support materials, i.e., nanoporous \ce{SiO2} beads and non-porous soda lime glass beads, at several applied voltages (10--20 kV) at near atmospheric pressure (550 Torr). Measurements of \ce{NH3} concentrations at the reactor outlet using a GC found that both the observed \ce{NH3} production rates and \ce{NH3} energy yields were enhanced in the presence of porous \ce{SiO2} beads compared to nonporous glass beads at the respective optimal \ce{N2}:\ce{H2} feed ratios for the two catalyst support materials. 

\par In conjunction with the \ce{NH3} synthesis measurements, we made operando measurements of Lissajous plots to characterize the plasma by the reduced electric field and electron number density, and measurements of high resolution OES spectra to obtain the relative concentrations of several active species \ce{N2+}, N, and H, in the plasma. OES measurements showed that the concentrations of \ce{N2+}, N, and H were lower for plasma in the presence of \ce{SiO2} than glass beads. Despite lower concentrations of \ce{N2+}, N, and H in the gas phase, the DBD reactor packed with \ce{SiO2} beads provided a higher \ce{NH3} yield than when packed with glass beads. This suggests that the active species in the gas phase were consumed more in the bed of \ce{SiO2} beads than glass beads, which is likely due to the fact that \ce{SiO2} provided larger surface area than glass. This shows that surface porosity of the catalyst support material has a significant effect on the plasma properties, reaction mechanism, and \ce{NH3} yield, and that surface reaction pathway contributes most significantly to the overall reaction of plasma-assisted \ce{NH3} synthesis. Higher surface porosity provides more surface sites for radical adsorption and the subsequent E--R reactions, resulting in higher overall reaction rate of plasma-assisted \ce{NH3} synthesis.

\section*{Acknowledgements}
The research described in this paper was, in part, supported by the Laboratory Directed Research and Development (LDRD) Program at Princeton Plasma Physics Laboratory, a national laboratory operated by Princeton University for the U.S. Department of Energy under Prime Contract No. DE-AC02-09CH11466. The United States Government retains a non-exclusive, paid-up, irrevocable, world-wide license to publish or reproduce the published form of this manuscript, or allow others to do so, for United States Government purposes. BEK acknowledges partial support of this work by the U.S. Department of Energy, Office of Science, Office of Fusion Energy Sciences under award number DE-SC0020233. ZC acknowledges partial support by the Program in Plasma Science and Technology (PPST) at Princeton University. SJ acknowledges Dr. Arthur Dogariu for providing the high resolution spectrometer and Dr. Evan Aguirre for helping with spectral analysis and reviewing the manuscript. The authors acknowledge Sonia Arumuganainar for the BET measurements. The authors acknowledge the use of Princeton's Imaging and Analysis Center (IAC), which is partially supported by the Princeton Center for Complex Materials (PCCM), a National Science Foundation (NSF) Materials Research Science and Engineering Center (MRSEC; DMR-2011750).

\section*{Data Availability}
The data that support the findings of this study are available from the corresponding author upon reasonable request.

%%%%%%%%%%%%%%%%%%%%%%
%\bibliography{sources.bib}

\begin{thebibliography}{34}%
\makeatletter
\providecommand \@ifxundefined [1]{%
 \@ifx{#1\undefined}
}%
\providecommand \@ifnum [1]{%
 \ifnum #1\expandafter \@firstoftwo
 \else \expandafter \@secondoftwo
 \fi
}%
\providecommand \@ifx [1]{%
 \ifx #1\expandafter \@firstoftwo
 \else \expandafter \@secondoftwo
 \fi
}%
\providecommand \natexlab [1]{#1}%
\providecommand \enquote  [1]{``#1''}%
\providecommand \bibnamefont  [1]{#1}%
\providecommand \bibfnamefont [1]{#1}%
\providecommand \citenamefont [1]{#1}%
\providecommand \href@noop [0]{\@secondoftwo}%
\providecommand \href [0]{\begingroup \@sanitize@url \@href}%
\providecommand \@href[1]{\@@startlink{#1}\@@href}%
\providecommand \@@href[1]{\endgroup#1\@@endlink}%
\providecommand \@sanitize@url [0]{\catcode `\\12\catcode `\$12\catcode
  `\&12\catcode `\#12\catcode `\^12\catcode `\_12\catcode `\%12\relax}%
\providecommand \@@startlink[1]{}%
\providecommand \@@endlink[0]{}%
\providecommand \url  [0]{\begingroup\@sanitize@url \@url }%
\providecommand \@url [1]{\endgroup\@href {#1}{\urlprefix }}%
\providecommand \urlprefix  [0]{URL }%
\providecommand \Eprint [0]{\href }%
\providecommand \doibase [0]{http://dx.doi.org/}%
\providecommand \selectlanguage [0]{\@gobble}%
\providecommand \bibinfo  [0]{\@secondoftwo}%
\providecommand \bibfield  [0]{\@secondoftwo}%
\providecommand \translation [1]{[#1]}%
\providecommand \BibitemOpen [0]{}%
\providecommand \bibitemStop [0]{}%
\providecommand \bibitemNoStop [0]{.\EOS\space}%
\providecommand \EOS [0]{\spacefactor3000\relax}%
\providecommand \BibitemShut  [1]{\csname bibitem#1\endcsname}%
\let\auto@bib@innerbib\@empty
%</preamble>
\bibitem [{\citenamefont {Mehta}\ \emph {et~al.}(2018)\citenamefont {Mehta},
  \citenamefont {Barboun}, \citenamefont {Herrera}, \citenamefont {Kim},
  \citenamefont {Rumbach}, \citenamefont {Go}, \citenamefont {Hicks},\ and\
  \citenamefont {Schneider}}]{1}%
  \BibitemOpen
  \bibfield  {author} {\bibinfo {author} {\bibfnamefont {P.}~\bibnamefont
  {Mehta}}, \bibinfo {author} {\bibfnamefont {P.}~\bibnamefont {Barboun}},
  \bibinfo {author} {\bibfnamefont {F.~A.}\ \bibnamefont {Herrera}}, \bibinfo
  {author} {\bibfnamefont {J.}~\bibnamefont {Kim}}, \bibinfo {author}
  {\bibfnamefont {P.}~\bibnamefont {Rumbach}}, \bibinfo {author} {\bibfnamefont
  {D.~B.}\ \bibnamefont {Go}}, \bibinfo {author} {\bibfnamefont {J.~C.}\
  \bibnamefont {Hicks}}, \ and\ \bibinfo {author} {\bibfnamefont {W.~F.}\
  \bibnamefont {Schneider}},\ }\href {\doibase
  https://doi.org/10.1038/s41929-018-0045-1} {\bibfield  {journal} {\bibinfo
  {journal} {Nature Catalysis}\ }\textbf {\bibinfo {volume} {\textbf{1}}},\
  \bibinfo {pages} {269–275} (\bibinfo {year} {2018})}\BibitemShut {NoStop}%
\bibitem [{\citenamefont {Bogaerts}\ and\ \citenamefont {Neyts}(2018)}]{2}%
  \BibitemOpen
  \bibfield  {author} {\bibinfo {author} {\bibfnamefont {A.}~\bibnamefont
  {Bogaerts}}\ and\ \bibinfo {author} {\bibfnamefont {E.~C.}\ \bibnamefont
  {Neyts}},\ }\href {\doibase https://doi.org/10.1021/acsenergylett.8b00184}
  {\bibfield  {journal} {\bibinfo  {journal} {ACS Energy Lett.}\ }\textbf
  {\bibinfo {volume} {\textbf{3}}},\ \bibinfo {pages} {1013–1027} (\bibinfo
  {year} {2018})}\BibitemShut {NoStop}%
\bibitem [{\citenamefont {Bogaerts}\ \emph {et~al.}(2020)\citenamefont
  {Bogaerts}, \citenamefont {Tu}, \citenamefont {Whitehead}, \citenamefont
  {Centi}, \citenamefont {Lefferts}, \citenamefont {Guaitella}, \citenamefont
  {Azzolina-Jury}, \citenamefont {Kim}, \citenamefont {Murphy}, \citenamefont
  {Schneider}, \citenamefont {Nozaki}, \citenamefont {Hicks}, \citenamefont
  {Rousseau}, \citenamefont {Thevenet}, \citenamefont {Khacef}, ,\ and\
  \citenamefont {Carreon}}]{1a}%
  \BibitemOpen
  \bibfield  {author} {\bibinfo {author} {\bibfnamefont {A.}~\bibnamefont
  {Bogaerts}}, \bibinfo {author} {\bibfnamefont {X.}~\bibnamefont {Tu}},
  \bibinfo {author} {\bibfnamefont {J.~C.}\ \bibnamefont {Whitehead}}, \bibinfo
  {author} {\bibfnamefont {G.}~\bibnamefont {Centi}}, \bibinfo {author}
  {\bibfnamefont {L.}~\bibnamefont {Lefferts}}, \bibinfo {author}
  {\bibfnamefont {O.}~\bibnamefont {Guaitella}}, \bibinfo {author}
  {\bibfnamefont {F.}~\bibnamefont {Azzolina-Jury}}, \bibinfo {author}
  {\bibfnamefont {H.-H.}\ \bibnamefont {Kim}}, \bibinfo {author} {\bibfnamefont
  {A.~B.}\ \bibnamefont {Murphy}}, \bibinfo {author} {\bibfnamefont {W.~F.}\
  \bibnamefont {Schneider}}, \bibinfo {author} {\bibfnamefont {T.}~\bibnamefont
  {Nozaki}}, \bibinfo {author} {\bibfnamefont {J.~C.}\ \bibnamefont {Hicks}},
  \bibinfo {author} {\bibfnamefont {A.}~\bibnamefont {Rousseau}}, \bibinfo
  {author} {\bibfnamefont {F.}~\bibnamefont {Thevenet}}, \bibinfo {author}
  {\bibfnamefont {A.}~\bibnamefont {Khacef}}, , \ and\ \bibinfo {author}
  {\bibfnamefont {M.}~\bibnamefont {Carreon}},\ }\href {\doibase
  https://iopscience.iop.org/article/10.1088/1361-6463/ab9048} {\bibfield
  {journal} {\bibinfo  {journal} {J. Phys. D: Appl. Phys.}\ }\textbf {\bibinfo
  {volume} {\textbf{53}}},\ \bibinfo {pages} {443001} (\bibinfo {year}
  {2020})}\BibitemShut {NoStop}%
\bibitem [{\citenamefont {Chen}\ \emph {et~al.}(2018)\citenamefont {Chen},
  \citenamefont {Crooks}, \citenamefont {Seefeldt}, \citenamefont {Bren},\ and\
  \citenamefont {et~al}}]{3}%
  \BibitemOpen
  \bibfield  {author} {\bibinfo {author} {\bibfnamefont {J.~G.}\ \bibnamefont
  {Chen}}, \bibinfo {author} {\bibfnamefont {R.~M.}\ \bibnamefont {Crooks}},
  \bibinfo {author} {\bibfnamefont {L.~C.}\ \bibnamefont {Seefeldt}}, \bibinfo
  {author} {\bibfnamefont {K.~L.}\ \bibnamefont {Bren}}, \ and\ \bibinfo
  {author} {\bibfnamefont {R.~M.~B.}\ \bibnamefont {et~al}},\ }\href {\doibase
  10.1126/science.aar6611} {\bibfield  {journal} {\bibinfo  {journal}
  {Science}\ }\textbf {\bibinfo {volume} {\textbf{360 (6391)}}} (\bibinfo
  {year} {2018}),\ 10.1126/science.aar6611}\BibitemShut {NoStop}%
\bibitem [{\citenamefont {et~al}(2020)}]{4}%
  \BibitemOpen
  \bibfield  {author} {\bibinfo {author} {\bibfnamefont {A.~B.}\ \bibnamefont
  {et~al}},\ }\href {\doibase
  https://iopscience.iop.org/article/10.1088/1361-6463/ab9048} {\bibfield
  {journal} {\bibinfo  {journal} {J. Phys. D: Appl. Phys}\ }\textbf {\bibinfo
  {volume} {\textbf{53}}},\ \bibinfo {pages} {443001} (\bibinfo {year}
  {2020})}\BibitemShut {NoStop}%
\bibitem [{\citenamefont {Hong}, \citenamefont {Prawer},\ and\ \citenamefont
  {Murphy}(2018)}]{5}%
  \BibitemOpen
  \bibfield  {author} {\bibinfo {author} {\bibfnamefont {J.}~\bibnamefont
  {Hong}}, \bibinfo {author} {\bibfnamefont {S.}~\bibnamefont {Prawer}}, \ and\
  \bibinfo {author} {\bibfnamefont {A.~B.}\ \bibnamefont {Murphy}},\ }\href
  {\doibase https://doi.org/10.1021/acssuschemeng.7b02381} {\bibfield
  {journal} {\bibinfo  {journal} {ACS Sustainable Chem. Eng.}\ }\textbf
  {\bibinfo {volume} {\textbf{6}}},\ \bibinfo {pages} {15--31} (\bibinfo {year}
  {2018})}\BibitemShut {NoStop}%
\bibitem [{\citenamefont {Sugiyama}\ \emph {et~al.}(1986)\citenamefont
  {Sugiyama}, \citenamefont {Akazawa}, \citenamefont {Oshima}, \citenamefont
  {Miura}, \citenamefont {Matsuda},\ and\ \citenamefont {Nomura}}]{6}%
  \BibitemOpen
  \bibfield  {author} {\bibinfo {author} {\bibfnamefont {K.}~\bibnamefont
  {Sugiyama}}, \bibinfo {author} {\bibfnamefont {K.}~\bibnamefont {Akazawa}},
  \bibinfo {author} {\bibfnamefont {M.}~\bibnamefont {Oshima}}, \bibinfo
  {author} {\bibfnamefont {H.}~\bibnamefont {Miura}}, \bibinfo {author}
  {\bibfnamefont {T.}~\bibnamefont {Matsuda}}, \ and\ \bibinfo {author}
  {\bibfnamefont {O.}~\bibnamefont {Nomura}},\ }\href {\doibase
  https://link.springer.com/article/10.1007/BF00571275} {\bibfield  {journal}
  {\bibinfo  {journal} {Plasma Chem. Plasma Process.}\ }\textbf {\bibinfo
  {volume} {\textbf{6}}},\ \bibinfo {pages} {179--193} (\bibinfo {year}
  {1986})}\BibitemShut {NoStop}%
\bibitem [{\citenamefont {Mizushima}\ \emph {et~al.}(2007)\citenamefont
  {Mizushima}, \citenamefont {Matsumoto}, \citenamefont {Ohkita},\ and\
  \citenamefont {Kakuta}}]{7}%
  \BibitemOpen
  \bibfield  {author} {\bibinfo {author} {\bibfnamefont {T.}~\bibnamefont
  {Mizushima}}, \bibinfo {author} {\bibfnamefont {K.}~\bibnamefont
  {Matsumoto}}, \bibinfo {author} {\bibfnamefont {H.}~\bibnamefont {Ohkita}}, \
  and\ \bibinfo {author} {\bibfnamefont {N.}~\bibnamefont {Kakuta}},\ }\href
  {\doibase https://link.springer.com/article/10.1007/s11090-006-9034-2}
  {\bibfield  {journal} {\bibinfo  {journal} {Plasma Chem. Plasma Process.}\
  }\textbf {\bibinfo {volume} {\textbf{27}}},\ \bibinfo {pages} {1--11}
  (\bibinfo {year} {2007})}\BibitemShut {NoStop}%
\bibitem [{\citenamefont {Akay}, ,\ and\ \citenamefont {Zhang}(2017)}]{8}%
  \BibitemOpen
  \bibfield  {author} {\bibinfo {author} {\bibfnamefont {G.}~\bibnamefont
  {Akay}}, , \ and\ \bibinfo {author} {\bibfnamefont {K.}~\bibnamefont
  {Zhang}},\ }\href {\doibase 10.1021/acs.iecr.6b02053} {\bibfield  {journal}
  {\bibinfo  {journal} {Ind. Eng. Chem. Res.}\ }\textbf {\bibinfo {volume}
  {\textbf{56(2)}}},\ \bibinfo {pages} {457–468} (\bibinfo {year}
  {2017})}\BibitemShut {NoStop}%
\bibitem [{\citenamefont {Hong}\ \emph {et~al.}(2017)\citenamefont {Hong},
  \citenamefont {Pancheshnyi}, \citenamefont {Tam}, \citenamefont {Lowke},
  \citenamefont {Prawer},\ and\ \citenamefont {Murphy}}]{9}%
  \BibitemOpen
  \bibfield  {author} {\bibinfo {author} {\bibfnamefont {J.}~\bibnamefont
  {Hong}}, \bibinfo {author} {\bibfnamefont {S.}~\bibnamefont {Pancheshnyi}},
  \bibinfo {author} {\bibfnamefont {E.}~\bibnamefont {Tam}}, \bibinfo {author}
  {\bibfnamefont {J.~J.}\ \bibnamefont {Lowke}}, \bibinfo {author}
  {\bibfnamefont {S.}~\bibnamefont {Prawer}}, \ and\ \bibinfo {author}
  {\bibfnamefont {A.~B.}\ \bibnamefont {Murphy}},\ }\href {\doibase
  https://iopscience.iop.org/article/10.1088/1361-6463/aa6229} {\bibfield
  {journal} {\bibinfo  {journal} {J. Phys. D: Appl. Phys.}\ }\textbf {\bibinfo
  {volume} {\textbf{50}}},\ \bibinfo {pages} {154005} (\bibinfo {year}
  {2017})}\BibitemShut {NoStop}%
\bibitem [{\citenamefont {Hong}\ \emph
  {et~al.}(2016{\natexlab{a}})\citenamefont {Hong}, \citenamefont {Aramesh},
  \citenamefont {Shimoni}, \citenamefont {Seo}, \citenamefont {Yick},
  \citenamefont {Greig}, \citenamefont {Charles}, \citenamefont {Prawer},\ and\
  \citenamefont {Murphy}}]{10}%
  \BibitemOpen
  \bibfield  {author} {\bibinfo {author} {\bibfnamefont {J.}~\bibnamefont
  {Hong}}, \bibinfo {author} {\bibfnamefont {M.}~\bibnamefont {Aramesh}},
  \bibinfo {author} {\bibfnamefont {O.}~\bibnamefont {Shimoni}}, \bibinfo
  {author} {\bibfnamefont {D.~H.}\ \bibnamefont {Seo}}, \bibinfo {author}
  {\bibfnamefont {S.}~\bibnamefont {Yick}}, \bibinfo {author} {\bibfnamefont
  {A.}~\bibnamefont {Greig}}, \bibinfo {author} {\bibfnamefont
  {C.}~\bibnamefont {Charles}}, \bibinfo {author} {\bibfnamefont
  {S.}~\bibnamefont {Prawer}}, \ and\ \bibinfo {author} {\bibfnamefont {A.~B.}\
  \bibnamefont {Murphy}},\ }\href {\doibase
  https://link.springer.com/article/10.1007/s11090-016-9711-8} {\bibfield
  {journal} {\bibinfo  {journal} {Plasma Chemistry and Plasma Processing}\
  }\textbf {\bibinfo {volume} {\textbf{50}}},\ \bibinfo {pages} {917–940}
  (\bibinfo {year} {2016}{\natexlab{a}})}\BibitemShut {NoStop}%
\bibitem [{\citenamefont {Zhu}\ \emph {et~al.}(2020)\citenamefont {Zhu},
  \citenamefont {Hu}, \citenamefont {Wu}, \citenamefont {Cai}, \citenamefont
  {Zhang},\ and\ \citenamefont {Tu}}]{11}%
  \BibitemOpen
  \bibfield  {author} {\bibinfo {author} {\bibfnamefont {X.}~\bibnamefont
  {Zhu}}, \bibinfo {author} {\bibfnamefont {X.}~\bibnamefont {Hu}}, \bibinfo
  {author} {\bibfnamefont {X.}~\bibnamefont {Wu}}, \bibinfo {author}
  {\bibfnamefont {Y.}~\bibnamefont {Cai}}, \bibinfo {author} {\bibfnamefont
  {H.}~\bibnamefont {Zhang}}, \ and\ \bibinfo {author} {\bibfnamefont
  {X.}~\bibnamefont {Tu}},\ }\href {\doibase
  https://iopscience.iop.org/article/10.1088/1361-6463/ab6cd1} {\bibfield
  {journal} {\bibinfo  {journal} {J. Phys. D: Appl. Phys.}\ }\textbf {\bibinfo
  {volume} {\textbf{53}}},\ \bibinfo {pages} {164002} (\bibinfo {year}
  {2020})}\BibitemShut {NoStop}%
\bibitem [{\citenamefont {Gómez-Ramírez}\ \emph {et~al.}(2015)\citenamefont
  {Gómez-Ramírez}, \citenamefont {Cotrino}, \citenamefont {Lambert}, ,\ and\
  \citenamefont {González-Elipe}}]{12}%
  \BibitemOpen
  \bibfield  {author} {\bibinfo {author} {\bibfnamefont {A.}~\bibnamefont
  {Gómez-Ramírez}}, \bibinfo {author} {\bibfnamefont {J.}~\bibnamefont
  {Cotrino}}, \bibinfo {author} {\bibfnamefont {R.~M.}\ \bibnamefont
  {Lambert}}, , \ and\ \bibinfo {author} {\bibfnamefont {A.~R.}\ \bibnamefont
  {González-Elipe}},\ }\href {\doibase
  https://iopscience.iop.org/article/10.1088/0963-0252/24/6/065011/pdf}
  {\bibfield  {journal} {\bibinfo  {journal} {Plasma Sources Sci. Technol.}\
  }\textbf {\bibinfo {volume} {\textbf{24}}},\ \bibinfo {pages} {065011}
  (\bibinfo {year} {2015})}\BibitemShut {NoStop}%
\bibitem [{\citenamefont {Gómez-Ramírez}\ \emph {et~al.}(2017)\citenamefont
  {Gómez-Ramírez}, \citenamefont {Montoro-Damas}, \citenamefont {Cotrino},
  \citenamefont {Lambert},\ and\ \citenamefont {González-Elipe}}]{13}%
  \BibitemOpen
  \bibfield  {author} {\bibinfo {author} {\bibfnamefont {A.}~\bibnamefont
  {Gómez-Ramírez}}, \bibinfo {author} {\bibfnamefont {A.~M.}\ \bibnamefont
  {Montoro-Damas}}, \bibinfo {author} {\bibfnamefont {J.}~\bibnamefont
  {Cotrino}}, \bibinfo {author} {\bibfnamefont {R.~M.}\ \bibnamefont
  {Lambert}}, \ and\ \bibinfo {author} {\bibfnamefont {A.~R.}\ \bibnamefont
  {González-Elipe}},\ }\href {\doibase https://doi.org/10.1002/ppap.201600081}
  {\bibfield  {journal} {\bibinfo  {journal} {Plasma Processes and Polymers}\
  }\textbf {\bibinfo {volume} {\textbf{14}}},\ \bibinfo {pages} {1600081}
  (\bibinfo {year} {2017})}\BibitemShut {NoStop}%
\bibitem [{\citenamefont {Neyts}\ \emph {et~al.}(2015)\citenamefont {Neyts},
  \citenamefont {Ostrikov}, \citenamefont {Sunkara},\ and\ \citenamefont
  {Bogaerts}}]{14}%
  \BibitemOpen
  \bibfield  {author} {\bibinfo {author} {\bibfnamefont {E.~C.}\ \bibnamefont
  {Neyts}}, \bibinfo {author} {\bibfnamefont {K.~K.}\ \bibnamefont {Ostrikov}},
  \bibinfo {author} {\bibfnamefont {M.~K.}\ \bibnamefont {Sunkara}}, \ and\
  \bibinfo {author} {\bibfnamefont {A.}~\bibnamefont {Bogaerts}},\ }\href
  {\doibase 10.1021/acs.chemrev.5b00362} {\bibfield  {journal} {\bibinfo
  {journal} {Chem. Rev.}\ }\textbf {\bibinfo {volume} {\textbf{115}}},\
  \bibinfo {pages} {13408--13446} (\bibinfo {year} {2015})}\BibitemShut
  {NoStop}%
\bibitem [{\citenamefont {Wang}\ \emph {et~al.}(2017)\citenamefont {Wang},
  \citenamefont {Yi}, \citenamefont {Wu}, \citenamefont {Guo},\ and\
  \citenamefont {Tu}}]{15}%
  \BibitemOpen
  \bibfield  {author} {\bibinfo {author} {\bibfnamefont {L.}~\bibnamefont
  {Wang}}, \bibinfo {author} {\bibfnamefont {Y.}~\bibnamefont {Yi}}, \bibinfo
  {author} {\bibfnamefont {C.}~\bibnamefont {Wu}}, \bibinfo {author}
  {\bibfnamefont {H.}~\bibnamefont {Guo}}, \ and\ \bibinfo {author}
  {\bibfnamefont {X.}~\bibnamefont {Tu}},\ }\href {\doibase
  https://doi.org/10.1002/anie.201707131} {\bibfield  {journal} {\bibinfo
  {journal} {Angew. Chem., Int. Ed.}\ }\textbf {\bibinfo {volume}
  {\textbf{56}}},\ \bibinfo {pages} {13679--13683} (\bibinfo {year}
  {2017})}\BibitemShut {NoStop}%
\bibitem [{\citenamefont {Wang}\ \emph {et~al.}(2018)\citenamefont {Wang},
  \citenamefont {Y.~Yi~Yi}, \citenamefont {Guo},\ and\ \citenamefont
  {Tu}}]{16}%
  \BibitemOpen
  \bibfield  {author} {\bibinfo {author} {\bibfnamefont {L.}~\bibnamefont
  {Wang}}, \bibinfo {author} {\bibfnamefont {Y.}~\bibnamefont {Y.~Yi~Yi}},
  \bibinfo {author} {\bibfnamefont {H.}~\bibnamefont {Guo}}, \ and\ \bibinfo
  {author} {\bibfnamefont {X.}~\bibnamefont {Tu}},\ }\href {\doibase
  https://doi.org/10.1021/acscatal.7b02733} {\bibfield  {journal} {\bibinfo
  {journal} {ACS Catal.}\ }\textbf {\bibinfo {volume} {\textbf{8}}},\ \bibinfo
  {pages} {90--100} (\bibinfo {year} {2018})}\BibitemShut {NoStop}%
\bibitem [{\citenamefont {Patil}(2017)}]{17}%
  \BibitemOpen
  \bibfield  {author} {\bibinfo {author} {\bibfnamefont {B.~S.}\ \bibnamefont
  {Patil}},\ }\emph {\bibinfo {title} {Plasma (Catalyst) Assisted Nitrogen
  Fixation: Reactor Development for Nitric Oxide and Ammonia Production}},\
  \href@noop {} {Ph.D. thesis},\ \bibinfo  {school} {Eindhoven University of
  Technology} (\bibinfo {year} {2017})\BibitemShut {NoStop}%
\bibitem [{\citenamefont {Herrera}\ \emph {et~al.}(2019)\citenamefont
  {Herrera}, , \citenamefont {Brown}, \citenamefont {abd Nazli~Turan},
  \citenamefont {Mehta}, \citenamefont {Schneider}, \citenamefont {Hicks},\
  and\ \citenamefont {Go}}]{18}%
  \BibitemOpen
  \bibfield  {author} {\bibinfo {author} {\bibfnamefont {F.~A.}\ \bibnamefont
  {Herrera}}, , \bibinfo {author} {\bibfnamefont {G.~H.}\ \bibnamefont
  {Brown}}, \bibinfo {author} {\bibfnamefont {P.~B.}\ \bibnamefont {abd
  Nazli~Turan}}, \bibinfo {author} {\bibfnamefont {P.}~\bibnamefont {Mehta}},
  \bibinfo {author} {\bibfnamefont {W.~F.}\ \bibnamefont {Schneider}}, \bibinfo
  {author} {\bibfnamefont {J.~C.}\ \bibnamefont {Hicks}}, \ and\ \bibinfo
  {author} {\bibfnamefont {D.~B.}\ \bibnamefont {Go}},\ }\href {\doibase
  https://iopscience.iop.org/article/10.1088/1361-6463/ab0c58} {\bibfield
  {journal} {\bibinfo  {journal} {J. Phys. D: Appl. Phys.}\ }\textbf {\bibinfo
  {volume} {\textbf{52}}},\ \bibinfo {pages} {224002} (\bibinfo {year}
  {2019})}\BibitemShut {NoStop}%
\bibitem [{\citenamefont {ul~Islam~Mujahid}\ \emph {et~al.}(2020)\citenamefont
  {ul~Islam~Mujahid}, \citenamefont {Kruszelnicki}, \citenamefont {Hala},\ and\
  \citenamefont {Kushner}}]{19}%
  \BibitemOpen
  \bibfield  {author} {\bibinfo {author} {\bibfnamefont {Z.}~\bibnamefont
  {ul~Islam~Mujahid}}, \bibinfo {author} {\bibfnamefont {J.}~\bibnamefont
  {Kruszelnicki}}, \bibinfo {author} {\bibfnamefont {A.}~\bibnamefont {Hala}},
  \ and\ \bibinfo {author} {\bibfnamefont {M.~J.}\ \bibnamefont {Kushner}},\
  }\href {\doibase https://doi.org/10.1016/j.cej.2019.123038} {\bibfield
  {journal} {\bibinfo  {journal} {Chemical Engineering Journal}\ }\textbf
  {\bibinfo {volume} {\textbf{382}}},\ \bibinfo {pages} {123038} (\bibinfo
  {year} {2020})}\BibitemShut {NoStop}%
\bibitem [{\citenamefont {Whitehead}(2019)}]{20}%
  \BibitemOpen
  \bibfield  {author} {\bibinfo {author} {\bibfnamefont {J.~C.}\ \bibnamefont
  {Whitehead}},\ }\href {\doibase https://doi.org/10.1007/s11705-019-1794-3}
  {\bibfield  {journal} {\bibinfo  {journal} {Frontiers of Chemical Science and
  Engineering}\ }\textbf {\bibinfo {volume} {\textbf{13}}},\ \bibinfo {pages}
  {264–273} (\bibinfo {year} {2019})}\BibitemShut {NoStop}%
\bibitem [{\citenamefont {Zhang}, \citenamefont {Neyts},\ and\ \citenamefont
  {Bogaerts}(2016)}]{21}%
  \BibitemOpen
  \bibfield  {author} {\bibinfo {author} {\bibfnamefont {Y.-R.}\ \bibnamefont
  {Zhang}}, \bibinfo {author} {\bibfnamefont {E.~C.}\ \bibnamefont {Neyts}}, \
  and\ \bibinfo {author} {\bibfnamefont {A.}~\bibnamefont {Bogaerts}},\ }\href
  {\doibase 10.1021/acs.jpcc.6b09038} {\bibfield  {journal} {\bibinfo
  {journal} {J. Phys. Chem. C}\ }\textbf {\bibinfo {volume} {\textbf{120}}},\
  \bibinfo {pages} {25923--25934} (\bibinfo {year} {2016})}\BibitemShut
  {NoStop}%
\bibitem [{\citenamefont {Barboun}\ \emph {et~al.}(2019)\citenamefont
  {Barboun}, \citenamefont {Mehta}, \citenamefont {Herrera}, \citenamefont
  {Go}, \citenamefont {Schneider},\ and\ \citenamefont {Hicks}}]{22}%
  \BibitemOpen
  \bibfield  {author} {\bibinfo {author} {\bibfnamefont {P.}~\bibnamefont
  {Barboun}}, \bibinfo {author} {\bibfnamefont {P.}~\bibnamefont {Mehta}},
  \bibinfo {author} {\bibfnamefont {F.~A.}\ \bibnamefont {Herrera}}, \bibinfo
  {author} {\bibfnamefont {D.~B.}\ \bibnamefont {Go}}, \bibinfo {author}
  {\bibfnamefont {W.~F.}\ \bibnamefont {Schneider}}, \ and\ \bibinfo {author}
  {\bibfnamefont {J.~C.}\ \bibnamefont {Hicks}},\ }\href {\doibase
  https://doi.org/10.1021/acssuschemeng.9b00406} {\bibfield  {journal}
  {\bibinfo  {journal} {ACS Sustainable Chem. Eng.}\ }\textbf {\bibinfo
  {volume} {\textbf{7} (9)}},\ \bibinfo {pages} {8621–8630} (\bibinfo {year}
  {2019})}\BibitemShut {NoStop}%
\bibitem [{\citenamefont {Mehta}\ \emph {et~al.}(2019)\citenamefont {Mehta},
  \citenamefont {Barboun}, \citenamefont {Go}, \citenamefont {Hicks}, ,\ and\
  \citenamefont {Schneider}}]{23}%
  \BibitemOpen
  \bibfield  {author} {\bibinfo {author} {\bibfnamefont {P.}~\bibnamefont
  {Mehta}}, \bibinfo {author} {\bibfnamefont {P.}~\bibnamefont {Barboun}},
  \bibinfo {author} {\bibfnamefont {D.~B.}\ \bibnamefont {Go}}, \bibinfo
  {author} {\bibfnamefont {J.~C.}\ \bibnamefont {Hicks}}, , \ and\ \bibinfo
  {author} {\bibfnamefont {W.~F.}\ \bibnamefont {Schneider}},\ }\href {\doibase
  https://doi.org/10.1021/acsenergylett.9b00263} {\bibfield  {journal}
  {\bibinfo  {journal} {ACS Energy Lett.}\ }\textbf {\bibinfo {volume}
  {\textbf{4} (5)}},\ \bibinfo {pages} {1115–1133} (\bibinfo {year}
  {2019})}\BibitemShut {NoStop}%
\bibitem [{\citenamefont {Gorky}, \citenamefont {Carreon},\ and\ \citenamefont
  {Carreon}(2020)}]{24}%
  \BibitemOpen
  \bibfield  {author} {\bibinfo {author} {\bibfnamefont {F.}~\bibnamefont
  {Gorky}}, \bibinfo {author} {\bibfnamefont {M.~A.}\ \bibnamefont {Carreon}},
  \ and\ \bibinfo {author} {\bibfnamefont {M.~L.}\ \bibnamefont {Carreon}},\
  }\href {\doibase https://doi.org/10.1088/2633-1357/aba1f8} {\bibfield
  {journal} {\bibinfo  {journal} {IOP SciNotes}\ }\textbf {\bibinfo {volume}
  {\textbf{1}}},\ \bibinfo {pages} {024801} (\bibinfo {year}
  {2020})}\BibitemShut {NoStop}%
\bibitem [{\citenamefont {Zhang}, \citenamefont {Neyts},\ and\ \citenamefont
  {Bogaerts}(2018)}]{25}%
  \BibitemOpen
  \bibfield  {author} {\bibinfo {author} {\bibfnamefont {Y.-R.}\ \bibnamefont
  {Zhang}}, \bibinfo {author} {\bibfnamefont {E.~C.}\ \bibnamefont {Neyts}}, \
  and\ \bibinfo {author} {\bibfnamefont {A.}~\bibnamefont {Bogaerts}},\ }\href
  {\doibase https://iopscience.iop.org/article/10.1088/1361-6595/aac0e4/meta}
  {\bibfield  {journal} {\bibinfo  {journal} {Plasma Sources Sci. Technol.}\
  }\textbf {\bibinfo {volume} {\textbf{27}}},\ \bibinfo {pages} {055008}
  (\bibinfo {year} {2018})}\BibitemShut {NoStop}%
\bibitem [{\citenamefont {Gu}\ \emph {et~al.}(2019)\citenamefont {Gu},
  \citenamefont {Zhang}, \citenamefont {Gao}, \citenamefont {Wang},
  \citenamefont {Zhang}, \citenamefont {Yi},\ and\ \citenamefont {Jiang}}]{26}%
  \BibitemOpen
  \bibfield  {author} {\bibinfo {author} {\bibfnamefont {J.-G.}\ \bibnamefont
  {Gu}}, \bibinfo {author} {\bibfnamefont {Y.}~\bibnamefont {Zhang}}, \bibinfo
  {author} {\bibfnamefont {M.-X.}\ \bibnamefont {Gao}}, \bibinfo {author}
  {\bibfnamefont {H.-Y.}\ \bibnamefont {Wang}}, \bibinfo {author}
  {\bibfnamefont {Q.-Z.}\ \bibnamefont {Zhang}}, \bibinfo {author}
  {\bibfnamefont {L.}~\bibnamefont {Yi}}, \ and\ \bibinfo {author}
  {\bibfnamefont {W.}~\bibnamefont {Jiang}},\ }\href {\doibase
  https://doi.org/10.1063/1.5082568} {\bibfield  {journal} {\bibinfo  {journal}
  {Journal of Applied Physics,}\ }\textbf {\bibinfo {volume} {\textbf{125}}},\
  \bibinfo {pages} {153303} (\bibinfo {year} {2019})}\BibitemShut {NoStop}%
\bibitem [{\citenamefont {Wang}\ \emph {et~al.}(2019)\citenamefont {Wang},
  \citenamefont {Craven}, \citenamefont {Yu}, \citenamefont {Ding},
  \citenamefont {Bryant}, \citenamefont {Huang},\ and\ \citenamefont
  {Tu}}]{27}%
  \BibitemOpen
  \bibfield  {author} {\bibinfo {author} {\bibfnamefont {Y.}~\bibnamefont
  {Wang}}, \bibinfo {author} {\bibfnamefont {M.}~\bibnamefont {Craven}},
  \bibinfo {author} {\bibfnamefont {X.}~\bibnamefont {Yu}}, \bibinfo {author}
  {\bibfnamefont {J.}~\bibnamefont {Ding}}, \bibinfo {author} {\bibfnamefont
  {P.}~\bibnamefont {Bryant}}, \bibinfo {author} {\bibfnamefont
  {J.}~\bibnamefont {Huang}}, \ and\ \bibinfo {author} {\bibfnamefont
  {X.}~\bibnamefont {Tu}},\ }\href {\doibase 10.1021/acscatal.9b02538}
  {\bibfield  {journal} {\bibinfo  {journal} {ACS Catal.}\ }\textbf {\bibinfo
  {volume} {\textbf{9}}},\ \bibinfo {pages} {10780--10793} (\bibinfo {year}
  {2019})}\BibitemShut {NoStop}%
\bibitem [{\citenamefont {Wagner}\ \emph {et~al.}(2003)\citenamefont {Wagner},
  \citenamefont {Brandenburg}, \citenamefont {Kozlov}, \citenamefont
  {Sonnenfeld}, \citenamefont {P.~Michel},\ and\ \citenamefont {Behnke}}]{28}%
  \BibitemOpen
  \bibfield  {author} {\bibinfo {author} {\bibfnamefont {H.~E.}\ \bibnamefont
  {Wagner}}, \bibinfo {author} {\bibfnamefont {R.}~\bibnamefont {Brandenburg}},
  \bibinfo {author} {\bibfnamefont {K.~V.}\ \bibnamefont {Kozlov}}, \bibinfo
  {author} {\bibfnamefont {A.}~\bibnamefont {Sonnenfeld}}, \bibinfo {author}
  {\bibfnamefont {P.}~\bibnamefont {P.~Michel}}, \ and\ \bibinfo {author}
  {\bibfnamefont {J.~F.}\ \bibnamefont {Behnke}},\ }\href {\doibase
  https://doi.org/10.1016/S0042-207X(02)00765-0} {\bibfield  {journal}
  {\bibinfo  {journal} {Vacuum}\ }\textbf {\bibinfo {volume} {\textbf{71}
  (3)}},\ \bibinfo {pages} {417--436} (\bibinfo {year} {2003})}\BibitemShut
  {NoStop}%
\bibitem [{\citenamefont {Hagelaar}\ and\ \citenamefont
  {Pitchford}(2005)}]{31}%
  \BibitemOpen
  \bibfield  {author} {\bibinfo {author} {\bibfnamefont {G.~J.~M.}\
  \bibnamefont {Hagelaar}}\ and\ \bibinfo {author} {\bibfnamefont {L.~C.}\
  \bibnamefont {Pitchford}},\ }\href {\doibase
  https://iopscience.iop.org/article/10.1088/0963-0252/14/4/011/meta}
  {\bibfield  {journal} {\bibinfo  {journal} {Plasma Sources Science and
  Technology}\ }\textbf {\bibinfo {volume} {\textbf{14} (4)}},\ \bibinfo
  {pages} {722} (\bibinfo {year} {2005})}\BibitemShut {NoStop}%
\bibitem [{\citenamefont {Butterworth}, \citenamefont {Elder},\ and\
  \citenamefont {Allen}(2016)}]{29}%
  \BibitemOpen
  \bibfield  {author} {\bibinfo {author} {\bibfnamefont {T.}~\bibnamefont
  {Butterworth}}, \bibinfo {author} {\bibfnamefont {R.}~\bibnamefont {Elder}},
  \ and\ \bibinfo {author} {\bibfnamefont {R.}~\bibnamefont {Allen}},\ }\href
  {\doibase https://doi.org/10.1016/j.cej.2016.02.047} {\bibfield  {journal}
  {\bibinfo  {journal} {Chemical Engineering Journal}\ }\textbf {\bibinfo
  {volume} {\textbf{293}}},\ \bibinfo {pages} {55--67} (\bibinfo {year}
  {2016})}\BibitemShut {NoStop}%
\bibitem [{\citenamefont {Mei}\ \emph {et~al.}(2014)\citenamefont {Mei},
  \citenamefont {Zhu}, \citenamefont {He}, \citenamefont {Yan},\ and\
  \citenamefont {Tu}}]{30}%
  \BibitemOpen
  \bibfield  {author} {\bibinfo {author} {\bibfnamefont {D.}~\bibnamefont
  {Mei}}, \bibinfo {author} {\bibfnamefont {X.}~\bibnamefont {Zhu}}, \bibinfo
  {author} {\bibfnamefont {Y.~L.}\ \bibnamefont {He}}, \bibinfo {author}
  {\bibfnamefont {J.~D.}\ \bibnamefont {Yan}}, \ and\ \bibinfo {author}
  {\bibfnamefont {X.}~\bibnamefont {Tu}},\ }\href {\doibase
  https://iopscience.iop.org/article/10.1088/0963-0252/24/1/015011} {\bibfield
  {journal} {\bibinfo  {journal} {Plasma Sources Science and Technology}\
  }\textbf {\bibinfo {volume} {\textbf{24} (1)}},\ \bibinfo {pages} {015011}
  (\bibinfo {year} {2014})}\BibitemShut {NoStop}%
\bibitem [{\citenamefont {Hong}\ \emph
  {et~al.}(2016{\natexlab{b}})\citenamefont {Hong}, \citenamefont {Aramesh},
  \citenamefont {Shimoni}, \citenamefont {Seo}, \citenamefont {Yick},
  \citenamefont {Greig}, \citenamefont {Charles}, \citenamefont {Prawer},\ and\
  \citenamefont {Murphy}}]{32}%
  \BibitemOpen
  \bibfield  {author} {\bibinfo {author} {\bibfnamefont {J.}~\bibnamefont
  {Hong}}, \bibinfo {author} {\bibfnamefont {M.}~\bibnamefont {Aramesh}},
  \bibinfo {author} {\bibfnamefont {O.}~\bibnamefont {Shimoni}}, \bibinfo
  {author} {\bibfnamefont {D.~H.}\ \bibnamefont {Seo}}, \bibinfo {author}
  {\bibfnamefont {S.}~\bibnamefont {Yick}}, \bibinfo {author} {\bibfnamefont
  {A.}~\bibnamefont {Greig}}, \bibinfo {author} {\bibfnamefont
  {C.}~\bibnamefont {Charles}}, \bibinfo {author} {\bibfnamefont
  {S.}~\bibnamefont {Prawer}}, \ and\ \bibinfo {author} {\bibfnamefont {A.~B.}\
  \bibnamefont {Murphy}},\ }\href {\doibase
  https://doi.org/10.1007/s11090-016-9711-8} {\bibfield  {journal} {\bibinfo
  {journal} {Plasma Chemistry and Plasma Processing}\ }\textbf {\bibinfo
  {volume} {\textbf{36}}},\ \bibinfo {pages} {917–940} (\bibinfo {year}
  {2016}{\natexlab{b}})}\BibitemShut {NoStop}%
\bibitem [{\citenamefont {Carrasco}\ \emph {et~al.}(2011)\citenamefont
  {Carrasco}, \citenamefont {Jiménez-Redondo}, \citenamefont {I.~Tanarro},\
  and\ \citenamefont {Herrero}}]{33}%
  \BibitemOpen
  \bibfield  {author} {\bibinfo {author} {\bibfnamefont {E.}~\bibnamefont
  {Carrasco}}, \bibinfo {author} {\bibfnamefont {M.}~\bibnamefont
  {Jiménez-Redondo}}, \bibinfo {author} {\bibfnamefont {I.}~\bibnamefont
  {I.~Tanarro}}, \ and\ \bibinfo {author} {\bibfnamefont {V.~J.}\ \bibnamefont
  {Herrero}},\ }\href {\doibase 10.1039/c1cp22284h} {\bibfield  {journal}
  {\bibinfo  {journal} {Physical Chemistry Chemical Physics,}\ }\textbf
  {\bibinfo {volume} {\textbf{13} (43)}},\ \bibinfo {pages} {19561--19572}
  (\bibinfo {year} {2011})}\BibitemShut {NoStop}%
\end{thebibliography}
%\end{thebibliography}%

%%%% Start Output BBL

\end{document}

\onecolumn
\section*{Supplementary Materials}

\begin{table}[h]
\centering
\begin{tabular}{|l|l|l|l|}
\hline
Material & Voltage (kV) & $E/N$ (Td) & $T_\text{e}$ (eV) \\
\hline \hline
\ce{SiO2} & 9 & 54 & 1.6 \\
& 11 & 61 & 2.0 \\
& 13 & 80 & 2.9 \\
& 15 & 94 & 3.4 \\
& 17 & 98 & 3.6 \\
& 19 & 110 & 4.0 \\
\hline \hline
glass & 11 & 61 & 2.0 \\
& 12 & 82 & 2.9 \\
& 13 & 86 & 3.1 \\
& 14 & 100 & 3.7 \\
& 15 & 117 & 4.3 \\
& 16 & 119 & 4.3 \\
& 17 & 126 & 4.5 \\
\hline
\end{tabular}
\caption{\label{tab:lines} Reduced electric field ($E/N$) and mean electron temperature ($T_\text{e}$) at different voltages for \ce{SiO2} and glass. Conditions: 550 Torr, 26.5 mL/min,  \ce{N2}:\ce{H2}=1:3. }
\end{table}